\documentclass[aps,twocolumn,groupedaddress,showpacs,floatfix]{revtex4}

\begin{document}
\bibliographystyle{apsrev}

\title{
Quantum ether: photons and electrons from a rotor model
}

\author{Michael Levin}
\author{Xiao-Gang Wen}
\homepage{http://dao.mit.edu/~wen}
\affiliation{Department of Physics, Massachusetts Institute of Technology,
Cambridge, Massachusetts 02139}
\date{May, 2005}

\begin{abstract} 
We give an example of a purely bosonic model -- a rotor model on the 3D cubic 
lattice -- whose low energy excitations behave like massless
$U(1)$ gauge bosons and 
massless Dirac fermions. This model can be viewed as a ``quantum ether'': a 
medium that gives rise to both photons and electrons. It illustrates a 
general mechanism for the emergence of gauge bosons and fermions known as 
``string-net condensation.'' Other, more complex, string-net condensed models 
can have excitations that behave like gluons, quarks and other particles in the
standard model. This suggests that photons, electrons and other elementary 
particles may have a unified origin: string-net condensation in our vacuum.
\end{abstract}

\maketitle

\section{Introduction}
Throughout history, people have attempted to understand the universe by 
dividing matter into smaller and smaller pieces. This approach has proven
extremely fruitful: successively smaller distance scales have revealed 
successively simpler and more fundamental structures. Over the last century, 
the fundamental building blocks of nature have been reduced from atoms to 
electrons, protons and neutrons, to most recently, the ``elementary'' particles
that make up the $U(1) \times SU(2) \times SU(3)$ standard model.
Today, a great deal of research is devoted to finding even more fundamental
building blocks - such as superstrings. 

This entire approach is based on the idea of reductionism - the idea 
that the fundamental nature of particles is revealed by dividing them into 
smaller pieces. But reductionism is not always useful or appropriate. For
example, in condensed matter physics there are particles, such as phonons,
that are collective excitations involving many atoms. These particles are 
``emergent phenomena'' that cannot be meaningfully divided into smaller pieces.
Instead, we understand them by finding the \emph{mechanism} that is
responsible for their emergence. In the case of phonons, for example, this 
mechanism is symmetry breaking.\cite{L3726,LanL58,N6080,G6154}

This suggests alternate line of inquiry. Could the elementary particles in the 
standard model be analogous to phonons? That is, could they be collective modes
of some ``structure'' that we mistake for empty space? 

Recent work suggests that they might be. \cite{LWuni,LWstrnet} This work has 
revealed the existence of new and exotic phases of matter whose collective 
excitations are gauge bosons and fermions. The microscopic degrees of freedom 
in these models are spins on a lattice - purely local, bosonic objects with 
local interactions. There is no trace of gauge boson or fermion degrees of 
freedom in the underlying lattice model. The gauge bosons and fermions are thus
emergent phenomena - a result of the collective behavior of many spins.

\begin{figure}[tb]
\centerline{
\includegraphics[scale=0.12]{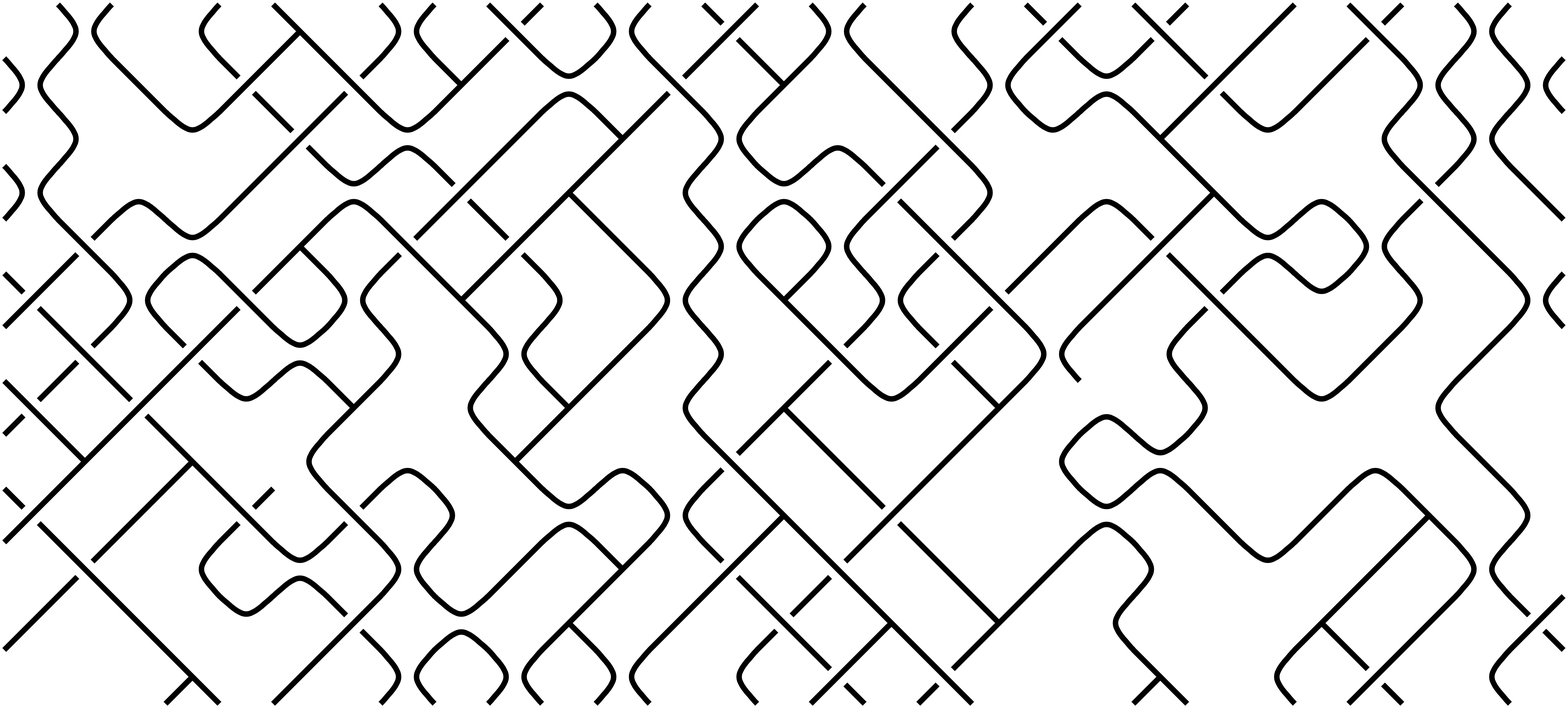}
}
\caption{
A picture of quantum ether: the fluctuations of the string-nets give rise to
gauge bosons (such as photons). The ends of the strings give rise to 
fermions (such as electrons).
}
\label{stringnetS}
\end{figure}

What is the mechanism responsible for their emergence? In these exotic phases,
the spins organize into a special pattern - a particular kind of entangled 
ground state, which we
call a ``string-net condensed'' state. A string-net condensed state is a spin
state where the spins organize into large string-like objects (or more
generally networks of strings). The strings then form a quantum string liquid
(see Fig. \ref{stringnetS}). This kind of ground state naturally gives rise to
gauge bosons and fermions. The gauge bosons correspond to fluctuations in the
strings - the collective motions of the strings that fill the space.
\cite{KS7595,BMK7793,Walight,Wen04}
The fermions correspond to endpoints of the strings - that is, defects in the
string liquid where a string ends in empty space. \cite{LWsta,LWuni,LWstrnet}

What makes the string-net picture particularly compelling is that the gauge
bosons and fermions naturally emerge \emph{together}. They are just different
aspects of the same underlying structure. Therefore, if we believe that the 
vacuum is such a string-net condensate then the presence of gauge interactions 
and Fermi statistics in the standard model is no longer mysterious. String-net 
condensation explains what gauge bosons and fermions are, why they exist, and 
why they appear together. \cite{LWuni}

The general theory of string-net condensation was worked out in \Ref{LWstrnet}.
One of the main results in that paper was a series of exactly soluble models 
realizing all possible string-net condensates. These models are quite general 
and can realize gauge bosons with any gauge group. However, they are also
complicated when discussed in full generality, and \Ref{LWstrnet} did not
provide an explicit example of the most physically relevant case - a model 
realizing gauge bosons and fermions in (3+1) dimensions.

In this paper, we attempt to remedy this problem. We demonstrate the string-net
picture of (3+1)D emerging gauge bosons and fermions with concrete lattice
models. We describe a rotor model on the cubic lattice that produces both
$U(1)$ gauge bosons and fermions.  The fermions can be gapped excitations (as
in an insulator) or gapless (as in a Fermi liquid). They can also behave like
massless Dirac fermions. In this case, the low energy physics of the rotor
model is identical to massless quantum electrodynamics (QED). The rotor model
can then be viewed as a ``quantum ether'': a medium that gives rise to both
photons and electrons.  In addition, the rotor model is closely related to
$U(1)$ lattice gauge theory coupled to a Higgs field. It demonstrates that a
simple modification or ``twist'' can change the Higgs boson into a fermion.

While this is not the first lattice bosonic model with emergent massless gauge 
bosons and massless Dirac fermions \cite{MA8938,Wlight,Wqoem,Wen04}, it has
two characteristics which distinguish it from previous examples. First,
the mapping between the rotor model and QED is essentially exact, and does not 
require a large $N$ limit or similar approximation. Second, the rotor model is 
a special case of a general construction, \cite{LWstrnet,LWuni}, unlike the 
other models which were in some sense, discovered by accident. It therefore 
provides a deeper understanding of emergent fermions and gauge bosons. In
addition to its relevance to high energy physics, this understanding may
prove useful to condensed matter physics, particularly the search for phases
of matter with exotic low energy behavior.

The paper is organized as follows: we begin with a ``warm-up'' calculation
in section II - a rotor model with emergent photons and \emph{bosonic} charges.
This model is closely related to $U(1)$ lattice gauge theory coupled to a
Higgs field. Then, in section III, we show that the rotor model can be 
modified in a natural way by adding a phase factor or ``twist'' to a term in 
the Hamiltonian. This modified or ``twisted'' rotor model has (massive) 
fermionic charges. In section IV, we describe further modifications which
give rise to gapless Fermi liquids and massless Dirac fermions.
  

\section{A 3D rotor model with emergent photons and bosonic charges}

In this section, we present a warm-up example - a rotor model with emergent 
photons and \emph{bosonic} charges. This model is closely related to $U(1)$ 
lattice gauge theory coupled to a Higgs field. While a number of other similar 
models \cite{MS0204,Walight,MS0312,HFB0404,AFF0493} have been analyzed 
previously, this model has the advantage of being quasi-exactly soluble, and 
generalizing easily to the fermionic case. The string-net picture of the ground
state is also particularly evident in this case.

\subsection{The model}

\begin{figure}[tb]
\centerline{
\includegraphics[scale=0.8]{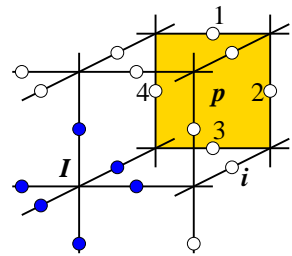}
}
\caption{
A picture of the rotor model (\ref{strnetH}). The term 
$Q_{\v I} = (-1)^{\v I}\sum_{\text{legs of }\v I} L_{\v i}^z$ acts on 
the six ``legs'' of $\v I$ - that is, the six rotors adjacent to $\v I$, drawn 
above as filled dots. The term $B_{\v p}= L^+_{1} L^-_{2} L^+_{3} L^-_{4}$ acts
on the four rotors, labeled by $1,2,3,4$, along the boundary of the plaquette 
$\v p$.
}
\label{cubbos}
\end{figure}

The model is a quantum rotor model with rotors on the links of a cubic lattice.
Each rotor can be viewed as a particle moving on a circle. The position of the 
particle is given by the angle $\th$, and the angular momentum of the particle 
by $L^z=I\dot\th$, where $I$ is the moment of inertia. The Hamiltonian of the 
rotor model is given by (see Fig. \ref{cubbos}a)\cite{MS0204,Walight}
\begin{align}
\label{strnetH}
H_\text{rotor} &= V\sum_{\v I}  Q^2_{\v I}
+ J \sum_{\v i} (L^z_{\v i})^2 - g \sum_{\v p} (B_{\v p}+h.c.)
\\
B_{\v p} &= L^+_{1}L^-_{2}L^+_{3}L^-_{4},\ \ \ \ \ \ \ \ \ 
Q_{\v I} = (-1)^{\v I} \sum_{\text{legs of }\v I} L^z_{\v i} ,
\nonumber 
\end{align}
where $\v I=(I_x, I_y, I_z)$ labels the vertices, $\v i$ labels the links and 
$\v p$ labels the plaquettes of the cubic lattice. The ``legs of $\v I$'' are 
the six links that are attached to the vertex $\v I$, and 1, 2, 3, 4 label the 
four links of the plaquette $\v p$  (see Fig. \ref{cubbos}). $L^\pm=e^{i\th}$ 
are the raising/lowering operators of $L^z$ and 
$(-1)^{\v I}\equiv (-1)^{I_x+I_y+I_z}$.

\subsection{The string-net picture}
\label{strnetpict}

\begin{figure}[tb]
\centerline{
\includegraphics[scale=0.8]{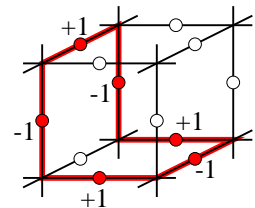}
}
\caption{
A picture of a closed string state. The rotors along the thick closed curve
have angular momentum $L^z = \pm 1$, while all the other rotors (depicted
as empty dots) have $L^z = 0$. 
}
\label{cubcstr}
\end{figure}

In this section, we show that the rotor model exhibits string-net condensation
in the regime $V \gg g \gg J > 0$. In particular, we show that the rotors
organize into effective extended objects, namely ``string-nets'', and the 
ground state is a quantum liquid of these string-nets. The string-net condensed
ground state has two types of excitations - string collective motions (which 
will correspond to photons) and string endpoints (which will correspond to 
charges).

The first step is to notice that the $Q_{\v I}$ operators commute with each 
other as well as the other terms in the Hamiltonian. Thus, we can label all the
eigenstates of $H_{\text{rotor}}$ by their eigenvalues under $Q_{\v I}$: 
$\{q_{\v I}\}$. The quantum number $q_{\v I}$ - which we will call the 
``charge'' on site $\v I$ - takes values in the integers.  

Clearly the lowest energy charge configuration is the charge-$0$ configuration 
$q_{\v I} = 0$. Other charge configurations cost an additional energy $V$. 
Hence, in the limit $V \gg J,g$, the low energy physics is completely contained
in the charge-$0$ sector.

We would like to enumerate the states in the charge-$0$ sector. It is natural 
to describe these states in terms of extended objects - in particular, 
string-nets. The simplest state has $L^z_{\v i}=0$ for every rotor. In the 
string-net language, we think of this state as the vacuum. Another state can be
obtained by alternately increasing or decreasing $L^z$ by $1$ along a loop 
(see Fig.~\ref{cubcstr}). We think of this state as the vacuum with a single 
(closed) string. Other states can be constructed by repeating this process. One
finds that the most general charge-$0$ state can have many strings and that the
strings can overlap to form networks of strings or ``string-nets.'' 

What is the action of the Hamiltonian (\ref{strnetH}) on these charge-$0$
states? If we think about the states using the string-net language, we see that
the $J$-term penalizes strings for being long. It therefore corresponds to
a string tension. On the other hand, the effect of the $B_{\v p}$ operator 
is to either create a small loop of small string (if applied to the vacuum) or
to deform the existing strings (if applied to a more complicated state).
Thus the $g$-term generates string ``hopping'' or string fluctuations. One
can think of it as a string kinetic energy.

There are two regimes to consider. When $J \gg g$, the string
tension dominates and the ground state contains almost no strings. The ground
state is a ``normal'' state. When $g \gg J$, the string kinetic energy 
dominates and the ground state is a superposition of many large closed strings 
\cite{Walight,Wen04}. The ground state is thus ``string-net condensed.'' We 
expect a quantum phase transition between these two states at some $J/g$ of 
order $1$.

We expect that the excitations above the ground state have very different 
properties in these two regimes. In the string-net condensed case, there are 
two types of excitations: low energy excitations with charge-$0$ and high 
energy excitations with nonzero charge.

The low energy excitations can be constructed from linear combinations of
string-net states. These excitations can be thought of as collective motions
of closed strings. 

The high energy, charged, excitations can be constructed from linear 
combinations of string-net configurations with \emph{open} strings. For 
example, if one takes the vacuum state $L^z = 0$ and alternately increases
and decreases $L^z$ along an open path $P$, the resulting state has nonzero
charge at the two ends of $P$. This is true quite generally: nonzero charge 
configurations are made up of open string-nets - with the charge located at 
the endpoints of the open strings. 

We can think of the string endpoints as point defects in the condensate. The 
higher energy excitations are composed out of these point defects. The defects 
carry a quantum number - charge - which is measured by the operator $Q_{\v I}$.

\subsection{Equivalence with $U(1)$ gauge theory}

In this section, we examine the rotor model more quantitatively. We show
that the low energy physics of the rotor model (\ref{strnetH}) is 
described by compact $U(1)$ gauge theory coupled to infinitely massive 
charges. The two types of excitations discussed in the previous section have 
a natural interpretation in terms of the gauge theory. The collective
motions of the strings 
give rise to gapless excitations and behave like photons, 
while the point defects (the endpoints of the strings) 
are gapped and behave like charged particles. 

In the gauge theory language, the strength of the interaction between the
photons (the gauge field) and the charges is characterized by a  ``fine
structure constant'' $\al$. The ``fine structure constant'' also characterizes
the strength of quantum fluctuations of the gauge field. When $\al\gg 1$, the
quantum fluctuations are so large that the gauge theory is in the
confining phase and there are no gauge bosons at low energies.
The gauge bosons (such as photons) can exist at low energies only
in the deconfined phase when $\al$ is small.

In the following, we will show that the ``fine structure constant'' $\al$ of
the gauge theory is of order $\sqrt{J/g}$. The ``speed of light'' $c$ is of
order $\sqrt{g Ja^2}$ where $a$ is the lattice constant.
This result provides insight into the phase transition between the normal and
string-net condensed states: the normal state corresponds to the large $\al$
confining phase of the gauge theory, while the string-net condensed state
corresponds to the small $\al$ deconfined phase. The string-net condensation
transition is the usual confinement-deconfinement transition from gauge theory.

The simplest way to derive these results is to map (\ref{strnetH}) onto a 
lattice gauge theory Hamiltonian. Consider the following Hamiltonian for
compact $U(1)$ gauge field coupled to a charge-$1$ Higgs field $e^{i\phi}$:
\begin{eqnarray}
\label{lgtH}
H_{\text{gauge}} &=& V \sum_{\v I} n_{\v I}^2 + 
J \sum_{\<\v{IJ}\>} E_{\v I \v J}^2 \nonumber \\ 
&-& 2g \sum_{\<\v{IJKL}\>} \cos(A_{\v I \v J}+A_{\v J \v K}+
A_{\v K \v L}+A_{\v L \v I}) \nonumber \\
&-& 2t \sum_{\<\v{IJ}\>} \cos(\phi_{\v I} - \phi_{\v J} - 
A_{\v I \v J}) 
\end{eqnarray}
Here, $\<\v{IJ}\>$, $\<\v{IJKL}\>$  is an alternative notation for links 
$\v i$, and plaquettes $\v p$ in the cubic lattice. The operator 
$E_{\v I \v J}$ is the integer valued electric field, canonically conjugate to 
the vector potential: $[A_{\v I \v J}, E_{\v I \v J}] = i$. Similarly, 
$n_{\v I}$ is the occupation number operator canonically conjugate to the Higgs
field: $[\phi_{\v I}, n_{\v I}] = i$. 

In lattice gauge theory, one is interested in the properties of this 
Hamiltonian within the gauge invariant subspace where Gauss' law holds:
\begin{equation}
\sum_{\<\v{IJ}\>} E_{\v I \v J} = n_{\v J}
\label{invsub}
\end{equation}
Our claim is that within this gauge invariant subspace, $H_{\text{gauge}}$ is
mathematically equivalent to $H_{\text{rotor}}$. To see this, note that 
the electric field operators and the occupation number operators
all commute with each other. So
the lattice gauge theory model has a complete basis consisting of states 
$|\{e_{\v I \v J}, n_{\v I}\}\>$ with electric field $e_{\v I \v J}$,
and occupation number $n_{\v I}$,
\begin{eqnarray*} 
E_{\v I \v J}|\{e_{\v I \v J}, n_{\v I}\}\> &=& 
e_{\v I \v J}|\{e_{\v I \v J}, n_{\v I}\}\> \\
n_{\v I}|\{e_{\v I \v J}, n_{\v I}\}\> &=& 
n_{\v I}|\{e_{\v I \v J}, n_{\v I}\}\>
\end{eqnarray*}
Similarly, the  rotor model $H_{\text{rotor}}$ has a complete basis consisting 
of states $|\{l^z_{\v i}\}\>$ with angular momentum $l^z_{\v I}$.

On can map the rotor model Hilbert space onto the gauge invariant subspace of 
the lattice gauge theory by mapping the basis state $|\{l^z_{\v i}\}\>$ onto 
the basis state $|\{e_{\v I \v J},n_{\v I}\}\>$ given by
\begin{eqnarray}
\label{lgtmap}
e_{\v I \v J} &=& (-1)^{\v I} \cdot l^z_{\v i} \nonumber \\
 n_{\v I} &=& (-1)^{\v I} \sum_{\text{legs of }\v I} l_{\v i}^z
\end{eqnarray}
(Here $\v i$ is the link connecting sites $\v I$ and $\v J$). Clearly, this 
correspondence maps the operators $n_{\v I}$ and $E_{\v I \v J}$ onto the 
operators $Q_{\v I}$ and $(-1)^{\v I} \cdot L^z_{\v I}$. One can also show that
this correspondence maps 
$\cos(A_{\v I \v J}+A_{\v J \v K}+A_{\v K \v L}+A_{\v L \v I})$ onto 
$1/2(B_{\v p} + h.c.)$ and 
$\cos(\phi_{\v I} - \phi_{\v J} - A_{\v I \v J})$ onto 
$\cos(\th_{\v i})$. Examining the two Hamiltonians, we conclude that the gauge 
theory \emph{in the special case $t =0$} can be mapped onto the rotor model.

The physics of the rotor model (\ref{strnetH}) is therefore completely 
equivalent to a $U(1)$ gauge field coupled to bosonic charges. The fine 
structure constant $\al$ and speed of light $c$ can be computed using the 
lattice gauge theory Hamiltonian (\ref{lgtH})). \cite{Walight,Wen04} As for
the mass of the charges, the rotor model (\ref{strnetH}) corresponds to the 
special case where $t = 0$ - the case where the charges are infinitely massive 
and have no dynamics. 

\subsection{Adding a charge hopping term}
In the previous section, we discovered that the point defect excitations 
have an infinite mass. This is not surprising, since the operator $Q_{\v I}$ 
commutes with the Hamiltonian (\ref{strnetH}), and hence charge configurations
are completely static.

We would like to modify $H_\text{rotor}$ so that the point defects 
(or ``charges'') do have dynamics. To do this, we need to add a term, 
$H_\text{hop}$, to the Hamiltonian that doesn't commute with $Q_{\v I}$. Such a
term will make the charges hop from site to site. We expect that any 
$H_\text{hop}$ will give rise to the small qualitative behavior (e.g. same 
particle statistics). We therefore consider the simplest hopping term:
\begin{equation*}
H_{\text{hop}} = -2t \sum_{\v{i}} \text{cos}({\th_{\v i}}) =
 -t \sum_{\v{i}} (L^+_{\v i} + L^-_{\v i})
\end{equation*}
It's easy to see that this term makes the charges hop from site to site just
as in a nearest neighbor tight binding model. In fact, if we examine the 
mapping, (\ref{lgtmap}), we see that $H_\text{hop}$ corresponds exactly with 
the nearest neighbor hopping term in lattice gauge theory, 
$\cos(\Delta_{k} \phi(\v r) - A_{k}(\v r))$. Therefore, the low energy physics 
of the rotor Hamiltonian $(H_\text{rotor} + H_\text{hop})$ is equivalent to 
that of a $U(1)$ gauge field coupled to charges with \emph{finite} mass. 

\subsection{The statistics of the $U(1)$ charges}
\label{statbos}
In this section, we compute the statistics of the charges and show that they 
are bosons. The main purpose of this computation is to facilitate comparison
with the ``twisted'' rotor model. (Indeed, the fact that the charges are bosons
follows immediately from the exact mapping to lattice gauge theory, 
\Eq{lgtmap}).

It is easiest to perform the computation in the case $J=0$. In this case,
$B_{\v p}$ commutes with the Hamiltonian $(H_{\text{rotor}} + H_{\text{hop}})$.
We can therefore divide the Hilbert space into different sectors 
$\{\th_{\v p}\}$ corresponding to different flux configurations 
$B_{\v p} = e^{i\th_{\v p}}$. A complete basis for each sector can be obtained 
by listing all the different charge configurations $\{q_{\v I}\}$. The action 
of the Hamiltonian $(H_{\text{rotor}} + H_{\text{hop}})$ on a charge 
configuration $\{q_{\v I}\}$ is simple. The first part of the Hamiltonian, 
$H_\text{rotor}$, doesn't affect the charge configuration at all, while the 
second part, $H_\text{hop}$ acts in two ways: it either creates two charges at 
neighboring sites, or it makes a charge hop from one site to another.

Thus, within each sector, $(H_\text{rotor}+H_\text{hop})$ is simply a hopping 
Hamiltonian on the cubic lattice. The hopping operators ${L}^{\pm}_{\v i}$ make
the charges hop from site $\v I$ to site $\v J$, where $\v I, \v J$ are the two
endpoints of $\v i$. To compute the statistics of the charges, we use the 
statistical hopping operator algebra described in \Ref{LWsta}. We note that the
hopping operators satisfy the algebra
\begin{equation}
L^{+}_{\v i}L^{-}_{\v j}L^{+}_{\v k} =
L^{+}_{\v k}L^{-}_{\v j}L^{+}_{\v i}
\end{equation}
for any $\v i,\v j, \v k$ incident to some vertex $\v I$. According to 
\Ref{LWsta}, this is the bosonic hopping operator algebra. We conclude that the
charged particles are bosons.

\section{A 3D rotor model with emergent photons and fermionic charges}
In this section, we present another rotor model which is very closely related 
to the model from the previous section, (\ref{strnetH}). The Hamiltonian only 
differs from $H_\text{rotor}$ by a ``twist'' - an additional phase factor in 
the $B_{\v p}$ term. Despite the apparent similarity, we will see that this 
``twisted'' rotor model gives rise to emergent photons and \emph{fermionic} 
charges. 

The relationship between the twisted and untwisted models is a special case of 
a general relationship between fermionic and bosonic string-net 
condensates. \cite{LWuni,LWstrnet} It is part of a systematic and 
general construction, unlike previous examples of emergent photons and fermions
\cite{Wlight,Wqoem,Wen04}, which were in some sense discovered by accident. 

\subsection{The twisted string operator}

\begin{figure}[tb]
\centerline{
\includegraphics[scale=.9]{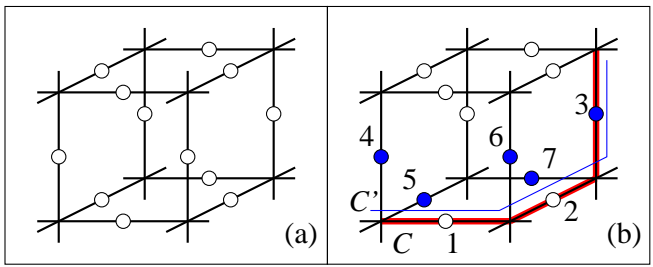}
}
\caption{
(a) A projection of the $3D$ cubic lattice onto the $2D$ plane.
(b) An example of a curve $C$ with a framing $C'$. The links along the string 
are labeled $1,2,3$. The links which cross $C'$ are the filled dots labeled by 
$3,4,5,6,7$. The corresponding twisted string operator $W^{\text{tw}}(C)$ is 
given by $L_1^+ L_2^- L_3^+ (-1)^{L^z_3+L^z_4+L^z_5 +L^z_6+L^z_7}$
}
\label{twstr}
\end{figure} 
In order to motivate the twisted rotor model, we first discuss the 
``twisted string operator.'' 

We begin with the original ``untwisted'' rotor model $H_{\text{rotor}}$. Notice
that the operator $B_{\v p} = L^+_1 L^-_2 L^+_3 L^-_4$ is a special case of a 
general string operator $W(C)$ that can be associated with any curve 
$C = \v i_1 \v i_2 \v i_3 ... \v i_n$ in the cubic lattice:
\begin{equation}
W(C) = L^+_{\v i_1}L^-_{\v i_2} L^+_{\v i_3}...L^-_{\v i_n}
\end{equation}
$B_{\v p}$ corresponds to the case where $C$ is the boundary
of a plaquette $\partial \v p$. 
 
This ``untwisted'' string operator has an important property. When $C$ is a 
closed curve, $W(C)$ commutes with all charge operators and all other closed 
string operators: 
\begin{equation}
\label{clstrid}
[Q_{\v I}, W(C)] =  [W(C'),W(C)] = 0
\end{equation}
This property is essential for the mapping to lattice gauge theory 
(\ref{lgtmap}) to hold. It implies that the $W(C)$ can be simultaneously 
diagonalized within a given charge sector. The simultaneous eigenstates can 
then be interpreted as different flux configurations, and the $W(C)$ can be 
interpreted as Wilson loop operators measuring the flux through a curve $C$.

It turns out that there is another string operator - which we call the 
``twisted string operator'' - that also has this property. The twisted 
string operator $\tilde{W}(C)$ is similar to the usual string operator 
$W(C)$ except that it contains an additional phase factor that depends on 
rotors on the ``legs'' of $C$ - that is, rotors on the links incident to 
vertices $\v I_k$ along the curve $C$.
 
To define $\tilde{W}(C)$, one must first choose a projection of
the $3D$ lattice onto a $2D$ plane. Any projection will work; in this paper, we
choose the particular projection illustrated in Fig. \ref{twstr}(a). 
The twisted string operator is then defined by
\begin{equation}
\label{twstrop}
\tilde{W}(C)=(L^+_{\v i_1} L^-_{\v i_2} ...) 
(-1)^{\sum_\text{legs of $C$} n^c_{\v i}L^z_{\v i}}
\end{equation}
Here, $n^c_{\v i}$ is the number of times that the link $\v i$ crosses the 
curve $C'$, where $C'$ is some ``framing'' of $C$. A ``framing'' is a 
curve $C'$ drawn next to $C$, but not exactly on it (see Fig. \ref{twstr}(b)). 
The choice of framing, $C'$, is not unique, so whenever we discuss 
$\tilde{W}$, we will need to specify a particular choice of framing.

One can check directly that both of the equalities in \Eq{clstrid} hold for the
twisted string $\tilde{W}$. Alternatively, one can refer to the general
argument given in \Ref{LWstrnet}.

\subsection{The twisted rotor model}

\begin{figure}[tb]
\centerline{
\includegraphics[scale=1.0]{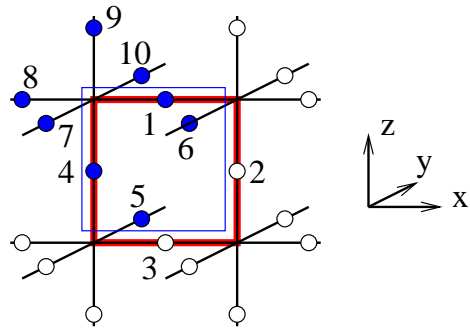}
}
\caption{
A picture of the $\tilde{B}_{\v p}$ term in the twisted rotor model 
(\ref{twistedH}). Just like any other twisted string operator, the term
$\tilde{B}_{\v p} = L_1^+ L_2^- L_3^+ L_4^- 
(-1)^{L_1^z+L_4^z+L_5^z+L_6^z+L_7^z+L_8^z+L_9^z+L_{10}^z}$ acts on the
four rotors, labeled by $1,2,3,4$, along the boundary of the plaquette $\v p$,
and the $8$ rotors, labeled by  $1,4,5,6,7,8,9,10$, that cross the framing 
curve.
}
\label{cubferm}
\end{figure}

The twisted model is obtained by modifying the $B_{\v p}$ term in 
(\ref{strnetH}). Instead of using a $B_{\v p}$ term based on the usual string 
operator, that is, $B_{\v p} = W(\partial \v p)$, we use a modified 
$B_{\v p}$ term based on the twisted string operator: 
$\tilde{B}_{\v p} = \tilde{W}(\partial \v p)$. We use a framing obtained
by taking the plaquette boundary $\partial \v p$ and shifting it up and to the
left (see Fig. \ref{cubferm}). The result is the twisted rotor Hamiltonian
\begin{align}
\label{twistedH}
&H^\text{tw}_\text{rotor} = V\sum_{\v I}  Q^2_{\v I}
+ J \sum_{\v i} (L^z_{\v i})^2 -  \sum_{\v p} g(\tilde{B}_{\v p}+h.c.)
\\
&Q_{\v I} = (-)^{\v I} \sum_{\text{legs of }\v I} L_{\v i}^z
\nonumber\\
&\tilde{B}_{\v p} 
= 
L_1^+ L_2^- L_3^+ L_4^- 
(-1)^{L_1^z+L_4^z+L_5^z+L_6^z+L_7^z+L_8^z+L_9^z+L_{10}^z}
\nonumber
\end{align}
where the above explicit definition of $\tilde{B}_{\v p}$ applies to 
plaquettes in the $xz$ plane (see Fig. \ref{cubferm}). (The definition for 
plaquettes in the $xy$ and $yz$ planes is similar and can be obtained using
$\tilde{B}_{\v p} = \tilde{W}(\partial \v p)$ together with the above 
framing convention).

Note that that (\ref{twistedH}) is very similar to the original rotor model
(\ref{strnetH}) (and hence also very similar to lattice gauge theory). 
The $\tilde{B}_{\v p}$ term alternately increases and decreases $L^z$ 
along the plaquette boundary $\partial \v p$, just like the $B_{\v p}$ term in
the model (\ref{strnetH}). The only difference is that here the amplitude for
this process is not always $1$. The amplitude is $\pm 1$, depending on the 
rotors on the legs of the plaquette boundary.

\subsection{Physical properties of the twisted rotor model}
The twisted rotor model can be analyzed in the same way as the untwisted model 
(Sec. \ref{strnetpict}). The charge operators $Q_{\v I}$ commute with the 
Hamiltonian, so all the states are associated with a charge configuration 
$\{q_{\v I}\}$. Low energy states have charge-$0$ and can be thought of as 
string-net states. They can be enumerated in the same way as in the untwisted 
model. Just as the untwisted closed-string states could be obtained by 
repeatedly applying different $B_{\v p}$'s to the vacuum state (the state with 
$L^z_{\v I} = 0$), the twisted closed-string states can be obtained by 
repeatedly applying different $\tilde{B}_{\v p}$'s to the vacuum state. Just as
before, the $J$ term can be interpreted as a string tension, and the $g$ term 
as a string kinetic energy. When $g \gg J$, string-net condensation occurs and 
the resulting ground state has two types of excitations: low energy string 
collective motions,
and higher energy defects in the condensate. 

More quantitatively, one can show that the low energy physics of the 
twisted rotor model is described by a compact $U(1)$ gauge field coupled to 
infinitely massive charges - just like the untwisted model. In fact, the 
twisted and untwisted rotor Hamiltonians (\ref{twistedH}), (\ref{strnetH}) are
completely equivalent and can be mapped onto one another.

To see this, note that the operators $Q_{\v I}$ and $\tilde{B}_{\v p}$ commute 
with each other and can therefore be simultaneously diagonalized. Let the 
simultaneous eigenstates be denoted $|\{q_{\v I}, \th_{\v p} \},\text{tw}\>$ 
where
\begin{eqnarray*}
Q_{\v I}|\{q_{\v I}, \th_{\v p} \},\text{tw}\> &=& 
q_{\v I}|\{q_{\v I}, \th_{\v p} \},\text{tw}\>
\\
\tilde{B}_{\v p}|\{q_{\v I}, \th_{\v p} \},\text{tw}\> &=& 
e^{i\th_{\v p}}|\{q_{\v I}, \th_{\v p} \},\text{tw}\>
\end{eqnarray*}
and the $q_{\v I}$, $\th_{\v p}$ are integers and real numbers, respectively.
The states $|\{q_{\v I}, \th_{\v p} \},\text{tw}\>$ form a complete basis
for the twisted rotor model. Similarly, we can simultaneously diagonalize
$Q_{\v I}$ and $B_{\v p}$, to construct a complete basis 
$|\{q_{\v I}, \th_{\v p} \}\>$ for the untwisted model. 

Note that the matrix elements of $\tilde{B}_{\v p}, Q_{\v I}$ in the twisted 
basis are the same as those of $B_{\v p}, Q_{\v I}$ in the untwisted basis. 
Furthermore the two sets of operators, 
$(Q_{\v I}, \tilde{B}_{\v p}, L^z_{\v i})$ and 
$(Q_{\v I},B_{\v p}, L^z_{\v i})$, satisfy the same commutation relations. 
Thus, $L^z_{\v I}$ must also have the same matrix elements in the two models. 
We conclude that the two Hamiltonians, (\ref{twistedH}) and (\ref{strnetH}), 
have exactly the same matrix elements (in their respective bases). The two
models are therefore equivalent.

\subsection{Adding charge dynamics}
As before, we would like to modify the Hamiltonian $H^\text{tw}_\text{rotor}$
so that the point defect excitations (or ``charges'') have a finite mass. 
To accomplish this, we need to add a term $H^\text{tw}_\text{hop}$ that 
doesn't commute with $Q_{\v I}$. Such a term will make the charges hop from 
site to site. 

We expect that any hopping term will give rise to the same qualitative 
behavior (e.g. same particle statistics). However, the resulting Hamiltonian is
easier to analyze if we pick a hopping term which commutes with 
$\tilde{B}_{\v p}$. While in the untwisted case the simplest hopping term 
happened to have this property, in the case at hand we have to work a little 
harder. 

\begin{figure}[t]
\centerline{
\includegraphics[scale=1.0]{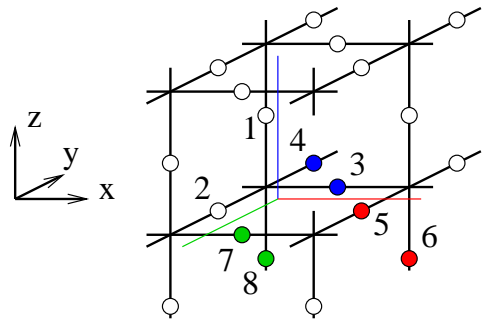}
}
\caption{
A picture of the hopping term 
$H^\text{tw}_\text{hop} = -
t \sum_{\v i} (\tilde{L}^+_{\v i}+\tilde{L}^-_{\v i})$. The 
$\tilde{L}^\pm_{\v i}$ are defined as twisted string operators 
$\tilde{W}(\v i)$, with the framing curve drawn just below and to the right
of $\v i$, as shown above. The result is
$\tilde{L}^\pm_{\v i} =L^{\pm}_{\v 3} (-1)^{L^z_5+L^z_6}$ when
$\v i$ is the link labeled $3$ pointing in the $\v x$ direction,
$\tilde{L}^\pm_{\v i}=L^{\pm}_{\v 2} (-1)^{L^z_7+L^z_8}$, when
$\v i$ is the link labeled $2$ pointing in the $\v y$ direction, and
$\tilde{L}^\pm_{\v i}=L^{\pm}_{\v 1} (-1)^{L^z_3+L^z_4}$ when
$\v i$ is the link labeled $1$ pointing in the $\v z$ direction.
}
\label{cubhop}
\end{figure}

We notice that the untwisted hopping term $H_\text{hop}$ is simply 
$-t(W(\v i) + W(\v i)^\dag)$ where $W(\v i)$ is the shortest open string 
operator (the string consists of a single link $\v i$). By analogy we choose 
the twisted hopping term to be $-t(\tilde{W}(\v i)+\tilde{W}(\v i)^\dag)$ 
with a framing convention given by drawing the framing curve just down and to
the right of the link $\v i$. The result is (see Fig. \ref{cubhop}):
\begin{equation}
H^\text{tw}_\text{hop} = -t \sum_{\v i} (\tilde{L}^+_{\v i}+
\tilde{L}^-_{\v i})
\end{equation}
where 
\begin{equation}
\tilde{L}^{\pm}_{\v i} = \left\{
\bmm
L^{\pm}_{\v i} (-1)^{L^z_5+L^z_6}, & 
\text{if} \ \ \v i \parallel \v x \\
L^{\pm}_{\v i} (-1)^{L^z_7+L^z_8}, & 
\text{if} \ \ \v i \parallel \v y \\
L^{\pm}_{\v i} (-1)^{L^z_3+L^z_4}, & 
\text{if} \ \ \v i \parallel \v z 
\emm
\right. 
\end{equation}
(One can check that $\tilde{L}^{\pm}_{\v i}$  does in fact commute with 
$\tilde{B}_{\v p}$ - either by explicit calculation or using the general 
properties of twisted string operators). \cite{LWstrnet}

Note that the term $H^\text{tw}_\text{hop}$ causes the charges to hop from 
sites to neighboring sites. The low energy physics of 
$(H^\text{tw}_\text{rotor} +H^\text{tw}_\text{hop})$ is therefore described by 
a compact $U(1)$ gauge field coupled to charges with \emph{finite} mass. 
However, these charges are not Higgs bosons as in the untwisted case. Indeed, 
with the addition of charge dynamics, the two Hamiltonians
$(H^\text{tw}_\text{rotor} +H^\text{tw}_\text{hop})$,
$(H_\text{rotor} +H_\text{hop})$, are no longer equivalent. The two models are 
only equivalent in the special case where $t=0$ and the charges have infinite 
mass. As soon as the charges can hop, the mapping between the two models breaks
down, as does the mapping to lattice gauge theory. The reason for this is that 
the charged particles in  $(H^\text{tw}_\text{rotor} +H^\text{tw}_\text{hop})$ 
are not bosons, but \emph{fermions} - as we will see in the next section.

\subsection{The statistics of the $U(1)$ charges}
In this section, we compute the statistics of the charges in the twisted rotor
model. We use the same technique as in Sec.\ref{statbos}.

As before, we perform the computation in the case $J=0$. Since 
$B^\text{tw}_{\v p}$ commutes with the Hamiltonian 
$(H^\text{tw}_{\text{rotor}} + H^\text{tw}_{\text{hop}})$, we can divide the 
Hilbert space into different sectors $\{\th_{\v p}\}$ corresponding to 
different flux configurations $B^\text{tw}_{\v p} = e^{i\th_{\v p}}$. 
The states within each sector can be labeled by their charge configuration 
$\{q_{\v I}\}$. The action of the Hamiltonian 
$(H^\text{tw}_{\text{rotor}} + H^\text{tw}_{\text{hop}})$ on a charge 
configuration $\{q_{\v I}\}$ is simple. The first part of the Hamiltonian, 
$H^\text{tw}_\text{rotor}$, doesn't affect the charge configuration at all, 
while the second part, $H^\text{tw}_\text{hop}$ acts in two ways: it either 
creates two charges at neighboring sites, or it makes a charge hop from one 
site to another.

Thus, within each sector, $(H^\text{tw}_\text{rotor}+H^\text{tw}_\text{hop})$ 
is simply a hopping Hamiltonian on the cubic lattice. The hopping operators 
$\tilde{L}^{\pm}_{\v i}$ make the charges hop from site $\v I$ to 
site $\v J$, where $\v I, \v J$ are the two endpoints of $\v i$. To compute the
statistics of the charges, we use the statistical hopping operator algebra
\cite{LWsta}, just as in the untwisted case. We find that the hopping
operators satisfy the same algebra but with a $-$ sign. That is,
\begin{equation}
\tilde{L}^{+}_{\v i}\tilde{L}^{-}_{\v j}\tilde{L}^{+}_{\v k} = (-1) \cdot
\tilde{L}^{+}_{\v k}\tilde{L}^{-}_{\v j}\tilde{L}^{+}_{\v i}
\end{equation}
for any $\v i,\v j, \v k$ incident to some vertex $\v I$. This sign makes
all the difference. The above algebra is not the bosonic hopping operator
algebra, but rather the \emph{fermionic} hopping algebra. We conclude that in
the twisted rotor model the charged particles are fermions.

One might expect that, by analogy with the untwisted case, there is a 
mapping to fermionic lattice gauge theory. However, no such mapping exists.
The reason is that, at the lattice scale, these fermions behave differently 
from the usual fermions in a tight binding model. 

There are two key differences. First, in a standard tight binding model of 
spinless fermions, each site can be occupied by $0$ or $1$ fermion. However,
in our case the charge $Q_{\v I}$ can be any integer from $-\infty$ 
to $\infty$. Thus, a given site can be occupied by arbitrarily many fermions. 
Second, in a standard tight binding model, the fermions are totally 
noninteracting in the zero coupling limit $J = 0$. However, in our case the
the fermions do interact, even in this limit. For example, the hopping
term $H_\text{hop}$ allows positive and negatively charged fermions to  
annihilate each other if they occupy neighboring sites. 

Neither of these differences should affect the physics of the twisted rotor 
model in the low energy limit, if $V$ is large. In that case, the low energy 
physics of the rotor model is equivalent to an insulator: it is described by 
two species of gapped fermions with opposite charges coupled to a $U(1)$ gauge 
field.

However, these microscopic differences could be important when $V$ becomes 
small enough that a Fermi surface develops. Unfortunately, in order to 
construct massless Dirac fermions, we need to consider that limit. Therefore, 
in order to make our calculation well-controlled, we need to modify the twisted
rotor model so that the fermions are equivalent to tight binding model 
fermions, \emph{even at the lattice scale}. This is the subject of the next 
section.

\section{Engineering massless Dirac fermions}
In this section, we modify the twisted rotor model 
$(H^\text{tw}_{\text{rotor}}+H^\text{tw}_{\text{hop}})$ so that the fermions
are exactly equivalent to tight binding model fermions. We then show that
this more well-controlled model can give rise to massless Dirac fermions.
It can also give rise to Fermi liquid behavior.

\subsection{The model}
There are many ways to make the fermions in 
$(H^\text{tw}_{\text{rotor}}+H^\text{tw}_{\text{hop}})$ behave like the usual
fermions from a tight binding model. Here we will only describe one of them. 
First, we divide the vertices of the cubic lattice into $3$ classes: 
``$eee$ vertices'', ``$ooo$ vertices'' and ``mixed vertices.'' A vertex is 
called an $eee$ vertex if the three components $\v I=(I_x,I_y,I_z)$ are all 
even, an $ooo$ vertex if the three components $\v I=(I_x,I_y,I_z)$ are all 
odd, and a mixed vertex otherwise. We then modify the $V$ term in 
(\ref{strnetH}) to
\begin{displaymath}
V \left( \sum_{eee} Q_{\v I} (Q_{\v I}-1) 
+ \sum_{ooo} Q_{\v I} (Q_{\v I}+1) 
+ \sum_{\text{mixed}} Q_{\v I}^2 \right) 
\end{displaymath}
Throughout our discussion we will take the $V\to +\infty$ limit. In this 
limit, $Q_{\v I}$ must be $0,\ 1$ on the $eee$ vertices, $0,\ -1$ on the $ooo$ 
vertices, and $0$ on the mixed vertices. In other words, the $U(1)$ charges can
only live on the $eee$ or $ooo$ vertices, and these two sublattices can only 
contain unit positive and negative charges respectively.

Next, we modify the hopping Hamiltonian $H^\text{tw}_{\text{hop}}$ to
\begin{equation}
\label{ffhop}
-t \sum_{\<\v i \v j\>, \text{coll.} } (\tilde{L}^+_{\v i}\tilde{L}^-_{\v j} + 
\tilde{L}^-_{\v i}\tilde{L}^{+}_{\v j})
\end{equation}
where the sum runs over pairs of neighboring links $\<\v i \v j\>$ which are 
collinear and can be joined together to form a single line segment of length 
$2$. This hopping term has the property that charges only move within the $eee$
and $ooo$ sublattices.

Putting this all together, and including a term which will correspond to a
chemical potential $\mu$, we arrive at the following rotor model:
\begin{align}
\label{Hff}
&H_\text{t.b} = 
J \sum_{\v i} (L^z_{\v i})^2 
- g\sum_{\v p} (\tilde{B}_{\v p}+h.c.) - 2\mu \sum_{\v i} L^z_{\v i}
\nonumber \\
& + V \left( \sum_{eee} Q_{\v I} (Q_{\v I}-1) 
+ \sum_{ooo} Q_{\v I} (Q_{\v I}+1) 
+ \sum_{\text{mixed}} Q_{\v I}^2 \right) 
\nonumber\\
&-t \sum_{\<\v i \v j\>, \text{coll.} } (\tilde{L}^+_{\v i}\tilde{L}^-_{\v j} +
\tilde{L}^-_{\v i}\tilde{L}^{+}_{\v j}))
\end{align}

\subsection{Solving the model}
We begin by considering the case where $t,J=0$. In this case $H_\text{t.b}$ is 
exactly soluble. Indeed, the basis states 
$|\{q_{\v I}, \th_{\v p} \}, \text{tw}\>$ are eigenstates of $H_\text{t.b}$
with an energy
\begin{align}
&E=  V \left( \sum_{eee} q_{\v I} (q_{\v I}-1) 
+ \sum_{ooo} q_{\v I} (q_{\v I}+1) 
+ \sum_{\text{mixed}} q_{\v I}^2 \right)
\nonumber \\
&-  \mu \sum_{\v I} (-1)^{\v I} q_{\v I} - 2g \sum_{\v p} \cos(\th_{\v p})
\label{Esolv}
\end{align}
As we mentioned earlier, we will take the $U\rightarrow \infty$ limit. In this 
limit, the only low energy states are those with $q_{\v I} = 0,1$ in the $eee$ 
sublattice, $q_{\v I}= 0,-1$ in the $ooo$ sublattice and $q_{\v I}=0$ 
everywhere else. If we restrict to this low energy subspace, the energy $E$
is simply given by
\begin{equation}
E =  -\mu N - 2g \sum_{\v p} \cos(\th_{\v p})
\end{equation}
where $N = \sum_{eee} q_{\v I} - \sum_{ooo} q_{\v I}$ is the total number
of charged particles. 

Now, consider the case $t \neq 0$. The $t$ term generates nearest neighbor 
hopping in the $eee$ and $ooo$ sublattices. If we let $P$ denote the projection
onto the low energy subspace, then the action of the hopping Hamiltonian 
(\ref{ffhop}) within this subspace can be written as
\begin{align*}
&\sum_{eee,\v \nu} 
(\tilde{t}_{\v I (\v I + 2\v \nu)} + h.c.) 
 +\sum_{ooo,\nu} 
(\tilde{t}_{\v I (\v I + 2\v \nu)} + h.c.) 
\end{align*}
where  $\v \nu$ runs over the unit vectors $\v x, \v y,\v z$ and the 
$\tilde{t}_{\v I (\v I+2\v \nu)}$ are hopping operators that make the 
particles hop from site $\v I + 2\v \nu$ to site $\v I$. They are defined by
\begin{equation}
\tilde{t}_{\v I(\v I+ \v 2\v \nu)} = 
-t \cdot P \tilde{L}^{\pm}_{\v i} \tilde{L}^{\mp}_{\v j} P 
\end{equation}
where $\v i$ is the link connecting $\v I, \v I + \v \nu$, $\v j$ is the link
connecting $\v I + \v \nu, \v I + \v 2\nu$, and we use the upper (lower) signs
for the $eee$ ($ooo$) sublattices. 

The low energy effective Hamiltonian can then be written as
\begin{eqnarray}
\label{Heff}
H_{\text{t.b, eff}} &=& \sum_{eee,\v \nu} (\tilde{t}_{\v I (\v I + 2\v \nu)} + 
h.c.)
\nonumber \\
 &+&\sum_{ooo,\v \nu} (\tilde{t}_{\v I (\v I + 2\v \nu)} + h.c.) \nonumber \\
&-& \mu N - 2g \sum_{\v p} \cos(\th_{\v p}) 
\end{eqnarray}
This Hamiltonian describes two types of hardcore charged particles with
charges $\pm 1$ hopping on two different sublattices. We can completely 
characterize this Hamiltonian by investigating two physical properties: the
statistics of the charges, and the gauge flux through each plaquette $\v p$.

To determine the statistics of the charges, we use the statistical hopping 
operator algebra. \cite{LWsta} We notice that the hopping operators satisfy the
relation
\begin{equation}
\label{ferhop}  
 \tilde{t}_{\v I \v L}
 \tilde{t}_{\v K \v I}
 \tilde{t}_{\v I \v J}
=
 (-1) \cdot
 \tilde{t}_{\v I \v J}
 \tilde{t}_{\v K \v I}
 \tilde{t}_{\v I \v L}
\end{equation}
for any $\v J,\v K,\v L$ adjacent to $\v I$. According to \cite{LWsta},
this is the fermionic hopping operator algebra, so the particles are fermions.

To determine the flux that the particles see, we compute the product
of hopping operators around a $2 \times 2$ plaquette $\<\v I \v J \v K \v L\>$ 
in the $eee$ or $ooo$ sublattice. We find that (up to some signs having to do 
with orientation conventions),
\begin{eqnarray}
\label{ferflux}
&&\tilde{t}_{\v I \v J}
\tilde{t}_{\v J \v K}
\tilde{t}_{\v K \v L}
\tilde{t}_{\v L \v I} \\
&=& (-t)^4 \cdot e^{\sum_{\v p \in 
\<\v I \v J \v K \v L\>} i \th_{\v p}}
n_{\v I} (1-n_{\v J})(1-n_{\v K})(1-n_{\v L})
\nonumber
\end{eqnarray}
Here the sum on the right hand side runs over the $4$ plaquettes in the 
original lattice that are contained in the (doubled) plaquette 
$\<\v I \v J \v K \v L\>$. Also, $n_{\v I} = (-1)^{\v I}Q_{\v I}$ denotes the 
occupation number of site $\v I$. This relation implies that the particles see 
a flux $\th_{\v p}$ through each plaquette $\v p$.

The hopping operators $\tilde{t}_{\v I \v J}$ are completely characterized by 
the two algebraic relations (\ref{ferhop}),(\ref{ferflux}). Any collection
of hardcore hopping operators that satisfy these relations will give rise to 
a Hamiltonian equivalent to (\ref{Heff}). A particularly simple collection of 
hopping operators satisfying these relations can be constructed from fermionic 
operators $c_{\v I}$ with
\begin{eqnarray*}
 \{c_{\v I}, c_{\v J} \} = 0, \ \ \
\{c_{\v I}, c^\dag_{\v J} \} = \del_{IJ} 
\end{eqnarray*}
The hopping operators are given by
\begin{equation}
\tilde{t}_{\v I (\v I + 2\v \nu)} = 
-t \cdot e^{\pm i (A_{\v I (\v I + \v \nu)} + 
A_{(\v I + \v \nu)(\v I + 2\v \nu)})}
c_{\v I}^\dag c_{\v I + 2\v \nu}
\end{equation}
where the upper (lower) sign applies to the $eee$ ($ooo$) sublattice and 
$e^{i A_{\v I \v J}}$ are phases defined on each link satisfying
\begin{displaymath}
e^{i A_{\v I_1\v I_n}}
e^{i A_{\v I_n\v I_{n-1}}}
\dots
e^{i A_{\v I_3\v I_2}}
e^{i A_{\v I_2\v I_1}} = e^{\sum_{\v p \in C} i \th_{\v p}}
\end{displaymath}
for any closed curve $C$. In other words, $A$ is a vector potential for the 
flux configuration $\th_{\v p}$. 

Rewriting the low energy effective Hamiltonian in terms of these
fermionic operators gives
\begin{eqnarray}
\label{Hfree}
&&H_{\text{t.b, eff}} \nonumber\\
&=& \sum_{eee,\v \nu} 
(-t \cdot e^{i (A_{\v I (\v I + \v \nu)} + 
A_{(\v I + \v \nu)(\v I + 2\v \nu)})}c_{\v I}^\dag c_{\v I + 2\v \nu}+ h.c.) 
\nonumber \\
&&+\sum_{ooo,\v \nu} 
(-t \cdot e^{-i (A_{\v I (\v I + \v \nu)} + 
A_{(\v I + \v \nu)(\v I + 2\v \nu)})}c_{\v I}^\dag c_{\v I + 2\v \nu}+ h.c.)
\nonumber \\
&&- \mu \sum_{eee} c^{\dag}_{\v I} c_{\v I}  - \mu 
\sum_{ooo} c^{\dag}_{\v I} c_{\v I} - 2g \sum_{\v p} \cos(\th_{\v p}) 
\end{eqnarray}
Note that this is nothing other than the standard tight binding Hamiltonian for
fermions hopping in fixed flux configurations $\th_{\v p}$. If we now 
allow $J \neq 0$, then the flux configurations acquire dynamics. The 
Hamiltonian (\ref{Hfree}) becomes modified by the addition of an electric 
energy term $J\sum_{\< \v I \v J \>} E^2_{\v I \v J}$, where $E_{\v I \v J}$ is
canonically conjugate to $A_{\v I \v J}$: $[A_{\v I \v J},E_{\v I \v J}] = i$.
This term is precisely what one needs for compact lattice $U(1)$ gauge theory. 
We conclude that in the general case $J,t \neq 0$, the twisted rotor model 
(\ref{Hff}) is mathematically equivalent to a tight binding model with
$\pm 1$ charged fermions coupled to a compact $U(1)$ gauge field.

\subsection{Fermi liquid behavior}
The Hamiltonian $H_\text{t.b,eff}$ naturally gives rise to Fermi liquid 
behavior. This is easiest to understand in the regime $g \gg |J|,|t|$. In this 
case, the minimum energy flux configuration is the configuration 
$\th_{\v p} = 0$. The flux $\th_{\v p}$ fluctuates about this minima giving 
rise to photon-like excitations. The fermion dispersion in the background flux 
configuration is given by
\begin{equation} 
E(\v k) = -2t(\cos(2k_x a) + \cos(2k_y a) +\cos(2k_z a)) - 
\mu 
\end{equation} 
The ground state is obtained by filling all the negative energy levels.
Clearly, the model behaves differently for different values of chemical 
potential $\mu$. If $\mu < - 6t$, then there are no negative energy levels.
All the energy levels are empty in the ground state, and the system is an 
insulator. In this case, the rotor model (\ref{Hff}) describes gapped 
fermions with a gap $\Delta = -6t - \mu$ coupled to a compact $U(1)$ gauge 
field. The fermions come in two species (corresponding to the two sublattices, 
$eee$ and $ooo$) with charges $+1$ and $-1$, respectively. Hence, just like the
original twisted rotor model 
$(H^\text{tw}_\text{rotor}+H^\text{tw}_\text{hop})$, the physics is equivalent
to that of a $U(1)$ gauge field coupled to two species of gapped fermions
with opposite charges.

On the other hand, if $\mu > -6t$, then the result is a Fermi liquid - or more 
precisely two Fermi liquids corresponding to the two species of fermions. These
two Fermi liquids are both coupled to a compact $U(1)$ gauge field.

\subsection{Massless Dirac fermions}
The model (\ref{Hfree}) can also give rise to massless Dirac fermions.
Indeed, massless Dirac fermions can occur whenever the fermion band structure 
has nodal points. This happens naturally in a $\pi$-flux configuration where 
each plaquette contains a flux of $\pi$. 

In our case, the relevant plaquettes are not plaquettes $\v p$ in the original 
cubic lattice, but rather ``doubled'' $2 \times 2$ plaquettes in the $eee$ and 
$ooo$ sublattices. One way to ensure that the flux through these ``doubled'' 
plaquettes is $\pi$, is to introduce a spatial dependence into the
parameter $t$. That is replace $t$ by $t \cdot \chi_{\v I \v J}$ where
\begin{align}
\label{U1aeee}
\chi_{\v I,\v I+2{\v x}} =& 1 , &
\chi_{\v I,\v I+2{\v y}} =& (-1)^{I_x/2},  \nonumber\\
\chi_{\v I,\v I+2{\v z}} =& (-1)^{(I_x+I_y)/2} 
\end{align}
on the $eee$ sublattice and
\begin{align}
\label{U1aooo}
\chi_{\v I,\v I+2{\v x}} =& 1 , &
\chi_{\v I,\v I+2{\v y}} =& (-1)^{(I_x-1)/2},  \nonumber\\
\chi_{\v I,\v I+2{\v z}} =& (-1)^{(I_x-1+I_y-1)/2} 
\end{align}
on the $ooo$ sublattice. The result is that \Eq{ferflux} becomes modified so 
that each fermion sees an effective flux of $\sum_{\v p \in 
\<\v I \v J \v K \v L\>} \th_{\v p} + \pi$ through each doubled plaquette
$\<\v I \v J \v K \v L\>$. 

The minimum energy flux configuration is still $\th_{\v p} = 0$, but now the 
fermions see an effective flux of $\pi$ through each doubled plaquette. The 
fermion dispersion in this background is given by $2$ pairs of degenerate bands
with energies
\begin{equation}
E(\v k) = \pm t \sqrt{ \cos^2(2k_x a) + \cos^2(2k_y a) + \cos^2(2k_z a)}-\mu 
\end{equation} 
and Brilluoin zone,
$(-\pi/4a, \pi/4a) \times (-\pi/4a, \pi/4a) \times (-\pi/2a, \pi/2a)$. The 
ground state of $H_\text{t.b, eff}$ is again obtained by filling all the levels
with negative energies. If $\mu = 0$, then the Fermi surface consists of $2$ 
nodal points with linear dispersion - at $\v k = (\pi/4a,\pi/4a,\pm \pi/4a)$. 
The low energy fermionic excitations are concentrated near these $2$ nodes. 
Together, the two nodes give rise to one four-component Dirac fermion in the 
continuum limit. Since the bands are doubly degenerate, and there are two 
species of fermions (corresponding to the $eee$ and $ooo$ sublattice), there 
are a total of $4$ massless four component Dirac fermions. 

We conclude that the low energy physics of the rotor model $H_\text{t.b}$ with 
$\mu = 0$, $V \rightarrow \infty$, $g \gg J,t$, and $t$ given by the above 
formula, is equivalent to QED with four species of massless Dirac fermions. 
Thus, $H_\text{t.b}$ is a local bosonic model that gives rise to both photons 
and massless fermions. It can be viewed as a realization of a quantum ether - 
an ether that produces not only light, but also electrons! 

We would like to point out that the speed of light $c\sim \sqrt{gJ} a$ and the
speed of the massless Dirac fermion $c_e \sim ta$ are not the same in our rotor
model. However, we can tune the value of $t$ to make $c=c_e$. In this case the
low energy effective theory will have Lorentz invariance. The reason that
we do not get a Lorentz invariant low energy effective theory naturally is
that our approach is based on Hamiltonian formalism where space and
time are treated very differently. If one uses a path integral formalism and
discretizes space and time in the same way, Lorentz invariance can
emerge naturally in a low energy effective theory. 

Another difficulty with the above model is that the emergent electrons are
massless. However, this problem can also be overcome:  when bosonic models
contain emergent non-Abelian gauge bosons, the models can produce electrons
with a \emph{finite} mass and such a mass can be much less than the cut-off
scale without fine tuning any parameters. \cite{Wqoem}

\section{Conclusion}

In this paper we have explicitly demonstrated the string-net picture of 
(3+1)D emerging gauge bosons and fermions. We presented a rotor model
on the cubic lattice with a string-net condensed ground state, and excitations
which behave just like massless photons and massless Dirac fermions. We saw 
that the model was closely related to $U(1)$ lattice gauge theory coupled to a 
Higgs field. The only real difference was a phase factor or ``twist'' in the 
Hamiltonian. Surprisingly, this ``twist''  was enough to turn the Higgs boson 
into a fermion.

This model is part of a much more general construction \cite{LWstrnet} that
can produce gauge bosons with any gauge group. In particular, one should be
able to construct a model that gives rise to $SU(3)$ gauge bosons and
massless Dirac fermions - that is, QCD. These models could have applications to
lattice gauge theory. They may make it possible to simulate QED with 
electrons and QCD with quarks without ever introducing quark or electron 
degrees of freedom on the sites. Only the link variables (representing the 
gauge field) would be necessary in such a simulation. 

From a high energy point of view, the string-net picture of the vacuum is
quite appealing. It explains why the standard model looks the way it does -
that is, why nature chooses such peculiar things as gauge bosons and fermions 
to describe itself. In addition, it unifies the mysterious gauge symmetries and
anticommuting fields into a single underlying structure: a string-net 
condensate.

But can we actually construct a string-condensed local bosonic model that 
produces the entire standard model? We are close, but not quite there. In terms
of elementary particles we can produce photons, gluons, leptons and quarks, but
we do not know how to produce neutrinos or $SU(2)$ gauge bosons.

The problem with the neutrinos and the $SU(2)$ gauge bosons is the famous
chiral-fermion problem. \cite{L0128} Neutrinos are chiral fermions and the 
$SU(2)$ gauge bosons couple chirally to other fermions. At the moment, we do 
not know how to obtain chiral fermions and chiral gauge theories from 
\emph{any} local lattice model, much less a local bosonic model.

From a condensed matter point of view, the above model shows how a simple spin 
system can give rise to emergent fermions and gauge bosons. Extended objects 
can give rise to both of these phenomena, as long as their condensate has an
appropriate ``twist.'' This may provide intuition in the search for new and 
exotic phases of matter - phases beyond the scope of Landay's theory of 
symmetry breaking. The discovery of a real material with string-net 
condensation would represent a breakthrough, particularly a material 
containing excitations which behave just like the photons and electrons in our 
vacuum.

This research is supported by NSF Grant No. DMR--04--33632 and by 
NSF-MRSEC Grant No. DMR--02--13282.



\end{document}